\documentclass[12pt]{article}
\usepackage{epsfig}
\newcommand{\ra}{\rightarrow}
\def\tgb{\tan \beta}
\def\be{\begin{equation}}
\def\ee{\end{equation}}
\newcommand{\LV}{\mbox{$\not \hspace{-0.09cm} L \hspace{0.05cm}$}}

\def\rp{R\!\!\!/ _p}
\begin{document}

\thispagestyle{empty}

\begin{center}
\begin{flushright}
                                                   CERN-TH/20002-305\\
                                                   hep-ph/0210425\\
\end{flushright}

\vspace{1cm}

{\Large

{\bf Three-body decays of $\tilde \chi_1^0$  in $\rp$ models
with dominant $\lambda$ and $\lambda'$ couplings \footnote{Talk presented
at the the 10th international conference on supersymmetry and unification 
of fundamental interactions, SUSY02, June 17-23, 2002, Hamburg (Germany).}
}}\\[5ex]
R.M. Godbole\footnote{On leave from Centre for
Theoretical Studies, Indian Institute of Science, Bangalore,
560~012, India. e-mail:rohini.godbole@cern.ch} \\ 
CERN, Theory Division, CH-1211, Geneva, Switzerland\\
\end{center}
\vspace{.5cm}
{\begin{center} ABSTRACT \end{center}}
\vspace{-4truemm}
{\small{
Decays of the lightest neutralino are studied in $R_p$-violating
models, with operators $\lambda^\prime L Q D^c$ and $\lambda L L E^c$
involving third-generation matter fields and with dominant
$\lambda^\prime$ and $\lambda$ couplings.  Decays with the
top-quark among the particles produced are considered, in addition to
those with an almost massless final state.  Phenomenological analyses
for examples of both classes of decays are presented.  No
specific assumption on the composition of the lightest neutralino is
made. Our formulae can easily be generalized to study
decays of heavier neutralinos. We also discuss effects of the $\rp$
decays of the $\tilde \chi_1^0$ on the possible faking of  the
$H^\pm$ signal at the LHC by the recently pointed out $\tilde \tau \bar b t$
production through $\lambda'_{333}$ coupling.}}

\newpage
\begin{center}
{\Large

{\bf Three-body decays of $\tilde \chi_1^0$  in $\rp$ models
with dominant $\lambda$ and $\lambda'$ couplings 
}}\\[5ex]
R.M. Godbole\footnote{On leave from Centre for
Theoretical Studies, Indian Institute of Science, Bangalore,
560~012, India. e-mail:rohini.godbole@cern.ch} \\
CERN, Theory Division, CH-1211, Geneva, Switzerland\\
\end{center}
\vspace{.5cm}
{\begin{center} ABSTRACT \end{center}}
\vspace{-4truemm}
{\small{
Decays of the lightest neutralino are studied in $R_p$-violating
models with operators $\lambda^\prime L Q D^c$ and $\lambda L L E^c$,
involving third-generation matter fields and with dominant
$\lambda^\prime$ and $\lambda$ couplings.  Decays with the
top-quark among the particles produced are considered, in addition to
those with an almost massless final state.  Phenomenological analyses
for examples of both classes of decays are presented.  No
specific assumption on the composition of the lightest neutralino is
made. Our formulae can easily be generalized to study
decays of heavier neutralinos. We also discuss effects of the $\rp$
decays of the $\tilde \chi_1^0$ on the possible faking of  the
$H^\pm$ signal at the LHC by the recently pointed out $\tilde \tau \bar b t$
production through $\lambda'_{333}$ coupling.}}


\section*{Introduction}
\label{intro} 

TeV scale Supersymmetry (SUSY) is theoretically the  most attractive way of 
stabilizing the mass of the Higgs boson against radiative corrections,
around the electroweak symmetry (EW)  breaking scale. Hence, 
SUSY searches, along with those for the Higgs boson, form the focus of physics 
studies at the current and  future colliders.
In the simplest version of SUSY models
one assumes conservation of a quantum number $R_p \equiv (-1)^{3B+L+2S}$, 
where $S$ stands for the spin of the particle. The conservation of this
quantum number is guaranteed only if the superpotential is invariant under
the discrete $R_p$  symmetry. However, there really is no deep theoretical 
reason for  this symmetry. As a matter of fact, it is 
possible to have $R_p$-violating terms in the superpotential while respecting
supersymmetry as well as invariance under the SM gauge transformation. These 
are given by 
\begin{equation}
W_{{\not{R}}_p}  =
\frac{1}{2} \lambda_{ijk} L_i L_j E^c_k +
            \lambda'_{ijk} L_i Q_j D^c_k +
\frac{1}{2} \lambda''_{ijk} U^c_i D^c_j D^c_k  +
            \kappa_i L_i H_2 .
\end{equation}
Here $L_i$,$Q_i$ are the doublet lepton, quark superfields, and
$E_i,U_i, D_i$ the singlet lepton, quark superfields, $i$ being the
generation index.  The above necessarily  mean non-conservation  of the 
baryon number $B$ and/or the lepton number $L$.  The $\LV$ couplings 
allow the generation of $\nu$ masses in an economical way without introducing 
any new fields. The contribution from the trilinear $\lambda,\lambda'$ 
comes at  the one- or two-loop level and the  bilinear $\kappa_i$  generate it
at tree level~\cite{fran1}. The Kamioka
and SNO experiments now give an unambiguious evidence for $\nu$ masses.
In SUSY models with $\rp$, there exists enough freedom to generate 
observed mass patterns with sparticle masses in the TeV range, and hence with
clean predictions for collider searches. This interplay between  
indirect constraints, coming from $\nu$ physics and  the collider signals 
of $\rp$ SUSY is a very interesting aspect of the $\rp$ SUSY studies.  
Of course the simultaneous existence of the 
$\lambda'$ and  $\lambda''$  couplings and TeV scale masses
for the sfermions will mean a very rapid proton decay. This can
be cured by adopting $B$ conservation, i.e. $\lambda'' = 0$. This choice 
is actually preferred if  $\rp$ terms are not to wash out baryogenesis.
Unified string theories actually prefer models with $B$ conservation and 
$\rp$~\cite{ross}. These models  treat the lepton and the quark fields 
differently and have two discrete symmetries. $B$ conservation and 
$\rp$ eliminate not just the dimension-$4$ operators for proton decay 
but also the dimension-$5$ operators. Needless to say that all 
this makes $\rp$ theoretically very interesting.

Of course the large
number, 48 as per Eq.~(1), of these essentially Yukawa-type couplings, with no
theoretical indications about their sizes, is  not a very satisfactory feature
of $\rp$ SUSY. Luckily many of these unknown couplings are constrained by low
energy processes such as $p/\mu$ decay, $\nu$ masses or
cosmological arguments like generating the right amount of baryon asymmetry in
the early Universe. There exist a host of constraints on $\rp$ 
couplings~\cite{us}. Most of the constraints come from virtual effects caused 
by sparticles in loops and some from direct collider searches. Many of these
get less severe with increasing number of the third-generation indices of the 
involved couplings. The constraints on the $\LV$ couplings coming from 
$\nu$ masses can be particularly severe but may also depend on model 
assumptions.  Hence a study of the {\it same} couplings  in  a collider 
environment,  can certainly help to clarify model building.

\section*{$R_p$ violation at the colliders}

Effects of $\rp$ at colliders fall into qualitatively different classes, 
depending  on the size of the $\rp$ couplings. The one effect that is 
always present is an unstable  lightest supersymmetric particle (LSP), 
which need not even be the $\tilde \chi_1^0$. If at least one of the 
$\rp$, $\lambda,\lambda'$ couplings is $ > 10^{-6}$, the LSP will decay 
in the detector.  In our work we consider the case of these $\LV, \rp$ 
couplings and take the  $\tilde \chi_1^0$ to be the LSP. The 
decays of the LSP $\tilde \chi_1^0 \rightarrow f \bar f_1 f_2 $
can give rise to strikingly different final states~\cite{meold} as 
compared to the SUSY signal with $R_p$ conservation.
For masses of $\tilde \chi_1^0, \tilde \chi_1^\pm$ of interest at
the LHC and NLC, even a $t$ in the final state will be allowed. For larger
$\rp$ couplings, 
decays of particles and sparticles other than the LSP via $\rp$ interactions
\cite{somedec,decon,dyus1} can take place. For example, $\tilde t \rightarrow b l , t \rightarrow b \tilde l$. These then may compete with the 
$R_p$-conserving decays. The $\rp$ couplings can also give contributions 
to  tree level processes due to virtual  sparticle exchange, such as
$pp \rightarrow \mu^+ \mu^-$~\cite{dyold,dyus1} or 
$ p p (p \bar p) \ra t \bar t X $.  In the last case the 
interference terms give rise to non-vanishing $t$ polarization~\cite{hik1}, 
which can be probed by the decay lepton energy/angular distribution~\cite{pou1}.
In addition, for large enough $\rp$ couplings, one also has resonant or 
non-resonant production of a single sparticle via the $\rp$ 
couplings~\cite{fran2,gia1,herbi}. For example, much as  $H^-$ produced 
via $t \bar b H^-$ coupling, one can also have
$p p \rightarrow t \bar b \tilde \tau$ via $\lambda'_{333}$ as shown in
\begin{figure}[hbt]
\centerline{\includegraphics[scale = 0.7]{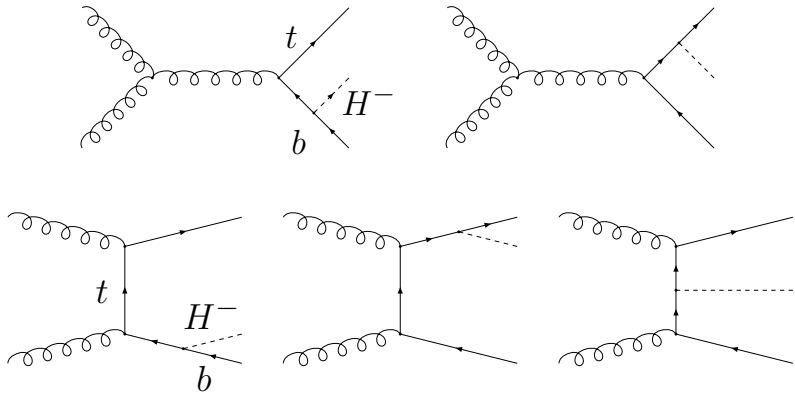}}
\caption{The different $2 \ra 3$ subprocesses contributing to 
$H^+/\tilde \tau$.}
\label{fig1}
\end{figure}
Figure~\ref{fig1}.  A few points are worth noting.  The production 
cross-section is
appreciable even, for $m_{\tilde t} > m_t$ and for $\lambda'_{333}$ 
as small as $0.05 $. $\tilde \tau$ so produced will have both $\rp$
and $R_p$-conserving decays. The latter will produce $\tilde \chi_1^0$,
which will further decay through some $\rp$ coupling.
Some of these decays can fake the charged Higgs signal.
This can be seen in a little more detail as follows.  For large 
$\lambda'_{333}, \tilde \tau '$s  can have the $R_p$ conserving decays, 
$\tilde \tau \rightarrow \tau \tilde \chi_1^0$ and  $\tilde \tau \rightarrow
\nu_\tau \tilde \chi_1^-$ for $m_t > m_{\tilde\tau}$, whereas for  
$m_t < m_{\tilde\tau}$, it can also have the $\rp$ decay 
$\tilde \tau \rightarrow b \bar t$. The 
net final state produced  will then be decided by
relative branching ratios of $\tilde \chi_1^0, \tilde \chi_1^+$ into
different channels. Production of $\tilde \tau$ through $\rp$ couplings 
and its decay via the {\it same} will give rise to
$p p \rightarrow t \bar b \tilde \tau X \rightarrow t \bar b t \bar b X$.
Note that this is the same final state as the $H^\pm$. 
A comprehensive study of the $H^+$ signal  at the LHC, in this case,
will thus require full analysis of the three-body decays of the 
$\tilde \chi_1^0, \tilde \chi_1^\pm$. The $R_p$-conserving 
decays of the $\tilde \tau$ followed by the $\rp$ decays of the
$\tilde \chi_1^0$ from the $\tilde \tau$ decay can also produce
\begin{eqnarray}
p p \rightarrow t \bar b \tilde \tau X &\rightarrow (2t) (2b) (2\tau) X
\nonumber \\
&\rightarrow tb~(2 \bar b)~\tau \nu_\tau X,
\end{eqnarray}
etc. These will then give rise to the characteristic $\LV$ 
signal of like-sign fermion pairs.

As mentioned already, of particular importance are the possibilities of 
testing the low energy constraints, in `direct' signals at the colliders.
Since the third generation sfermions are expected to be lighter,
they may give rise to larger virtual effects.  For $\tilde \chi_1^0, 
\tilde \chi_1^+$ with masses of interest at the LHC, NLC,  final states 
with third-generation fermions including $t$ are possible. One needs a study 
of the $\rp$ decays of $\tilde \chi_1^0, \tilde \chi_1^\pm$ retaining 
effects of the mass of the third-generation fermions, for $\LV$ coupling.

We therefore calculated three-body $\rp$ decays of the $\tilde \chi_1^0$
with dominant $\lambda, \lambda'$ couplings involving at least two 
third-generation indices.  We  kept the mass effects of the third-generation 
fermions and  the sfermion $L$--$R$ mixing terms complex and analysed
numerically for cases with unified/ununified gaugino masses,
including the effect of subdominant $\lambda,\lambda'$. 

\section*{Three-body decays of the $\tilde\chi_1^0$}
In this section we present the results for the above-mentioned
case of dominant $\lambda'_{333}$ coupling. The calculation and the results 
can be trivially generalized~\cite{us} to cases of other couplings, dominant
or subdominant. 
\begin{figure}
\centerline{\epsfig{file=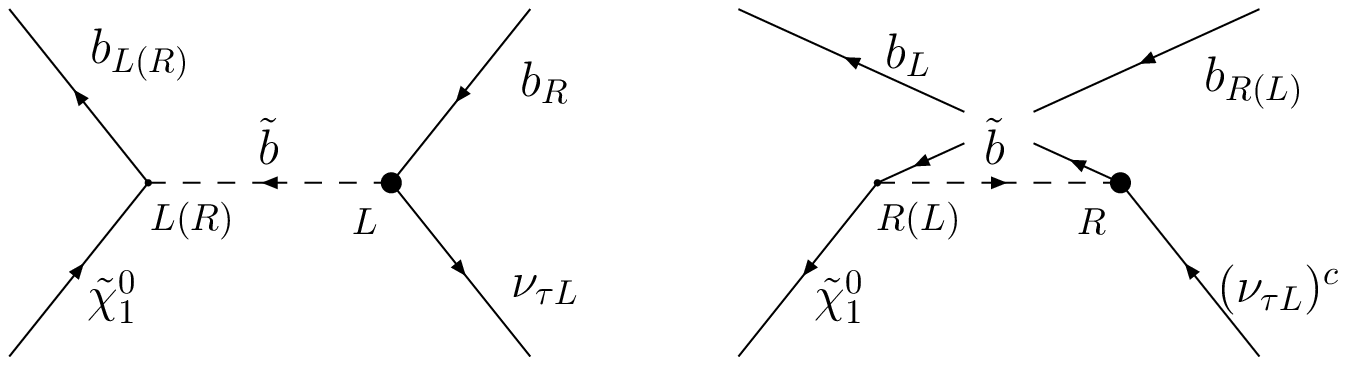,height=4cm,width = 12cm}
\epsfig{file=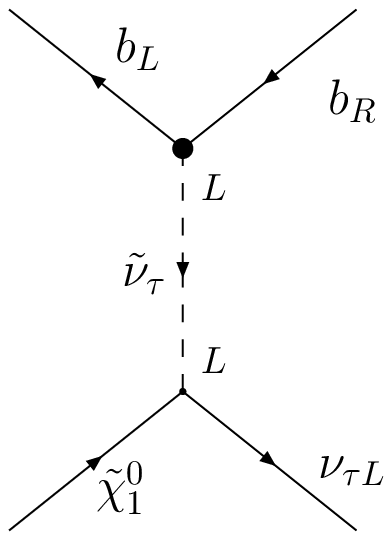,height=4cm}}
\caption{Diagrams contributing to the three body decay 
$\tilde \chi_1^0 \rightarrow b \bar b \nu_\tau$.}
\label{fig2}
\end{figure}
Figure~\ref{fig2} shows the diagrams contributing to the three-body decay 
$\tilde \chi_1^0 \rightarrow b \bar b \nu_\tau$. If 
$m_{\tilde \chi_1^0} > m_t$, then the decay into a massive final state 
containing
$t$, viz. $\tilde \chi_1^0 \rightarrow \bar b t \nu_\tau$ + C.C., is also 
possible. Our expressions for the decay widths~\cite{us} are the most 
general in that they include the effect of the finite fermion masses and 
are applicable to the case of complex $L$--$R$ mass term. In the limit
of real $L$--$R$ sfermion mass terms, these agree with the earlier 
calculations~\cite{herb2} with finite fermion mass and, in the limit of 
zero fermion mass, they agree with those of Refs.~\cite{gond1,us1} up to the 
corrections pointed out in Ref.~\cite{herb2}. 

The decay widths clearly depend on the $L$--$R$ mixing in the sfermion 
sector as well as on the composition of the $\tilde \chi_1^0$. For example,
for a Wino-like $\tilde \chi_1^0$, in the absence of $L$--$R$ mixing, the 
second diagram with a $\tilde {b}_R$ exchange will not contribute.
A substantial gaugino/higgsino mixing causes the width into the massive mode 
to be large at low $\tan \beta$, whereas for large $\tan \beta$ it is
the massless mode that gets enhanced, compared to the corresponding
widths for unmixed cases. 

As an illustration, the results for the $\tilde\chi_1^0$ widths into massive
and massless modes, for $\lambda'_{333} = 1$ and $M_{\tilde {\chi_1^0}} =  
600 $  GeV, are presented in
\begin{figure}[hbt]
\includegraphics*[scale=0.35]{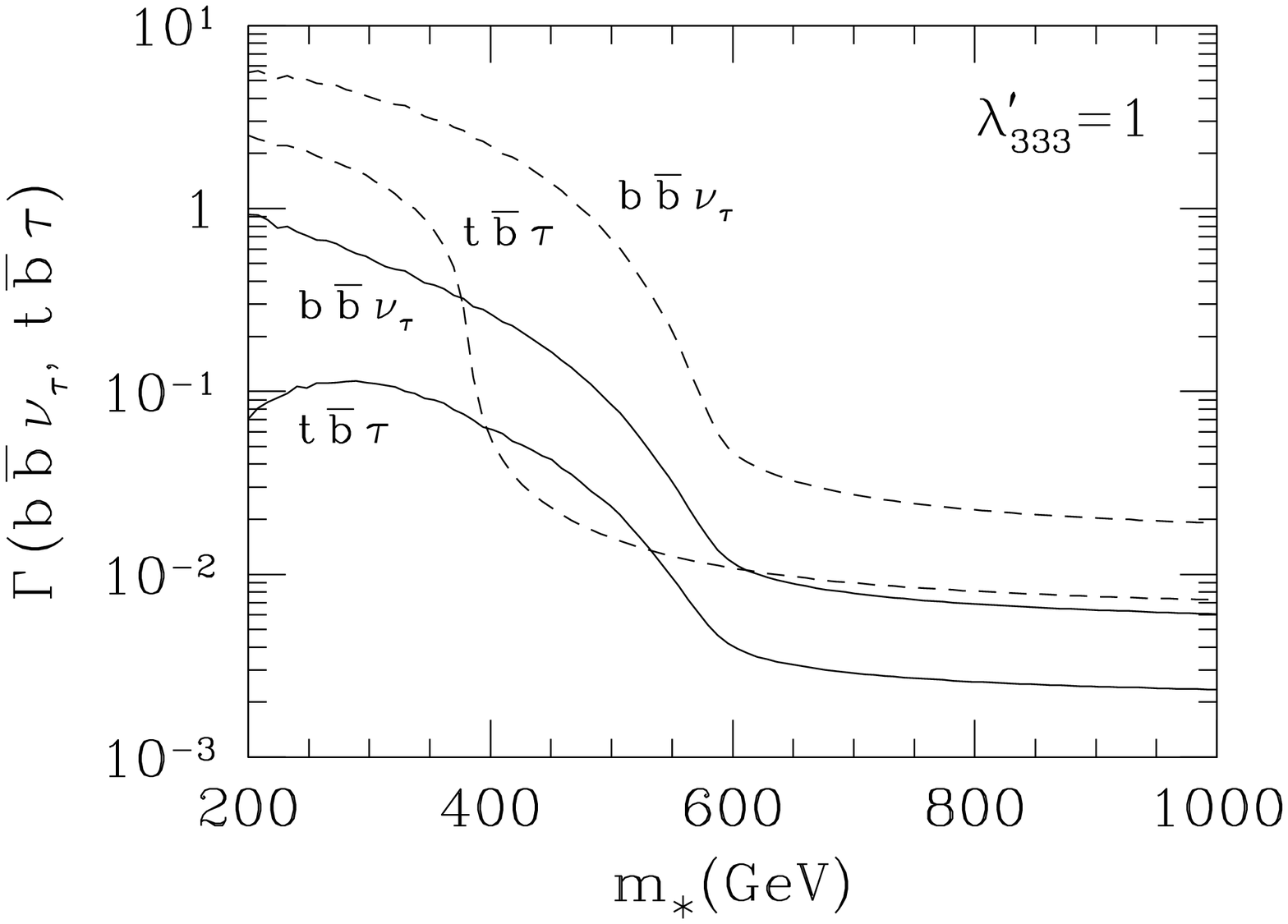}
\includegraphics*[scale=0.35]{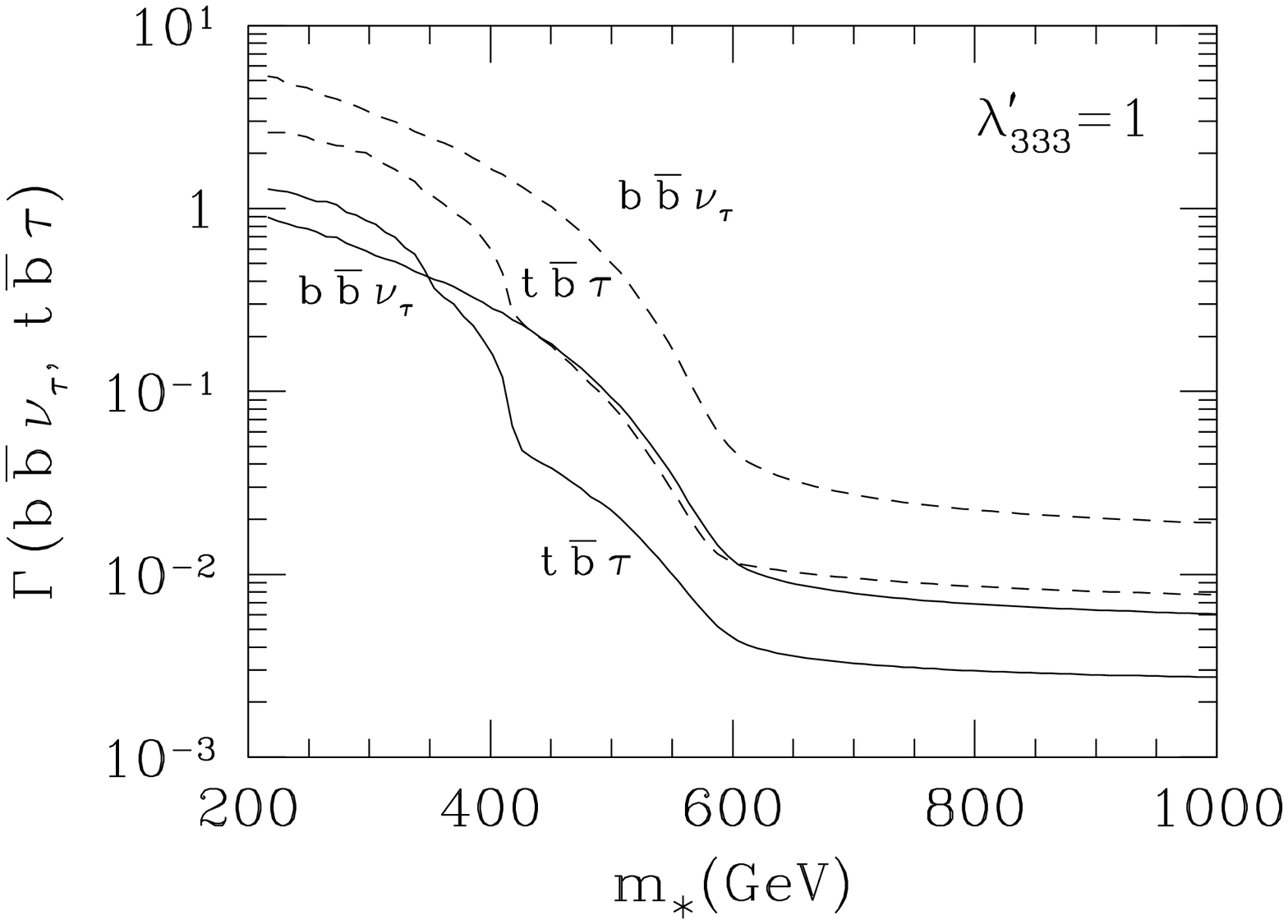}
\caption{Left panel shows $\tilde \chi_1^0$ widths for no $L$--$R$ squark 
mixing and the right one for moderate mixing. Dashed and solid lines are
for a Wino-like  and  Bino-like $\tilde \chi_1^0$ respectively. In case of
non-zero mixing, $A_t - \mu \cot \beta = 150$ GeV, $A_b - \mu \tan \beta = 
2000$ GeV.  $m_*$ is a squark  mass scale.}
\label{fig3}
\end{figure}
Figure~\ref{fig3} as a function of  $m_*$, the  squark  mass scale.
We use  two different choices of $L$--$R$ mixing in
the squark sector and different compositions of the $\tilde\chi_1^0$.  
The massive decay mode has a large width for low 
$\tan \beta$ and large higgsino--gaugino mixing.
Otherwise the massless mode is larger, although the massive mode is
non-negligible.  From the left panel  we see that, for a
$\tilde {\chi_1^0} \sim \tilde W$, the widths are enhanced by an 
order of magnitude with respect to the case of a 
$\tilde \chi_1^0 \sim \tilde B$. The right panel shows that
with moderate $L$--$R$ squark mixing, even for a
$\tilde \chi_1^0 \sim \tilde W,$ the $\tilde b$-mediated diagram
contributes to the massive mode. Further the figures also show that
smaller $\tilde {t_1}, \tilde {b_1}$, masses enhance the width for the 
massive mode for a $\tilde B$-like $\tilde \chi_1^0$. For the same reasons the
widths into the massive and massless final state also depend on the value of 
$\tgb$. 

\begin{figure}
\includegraphics*[scale=0.35]{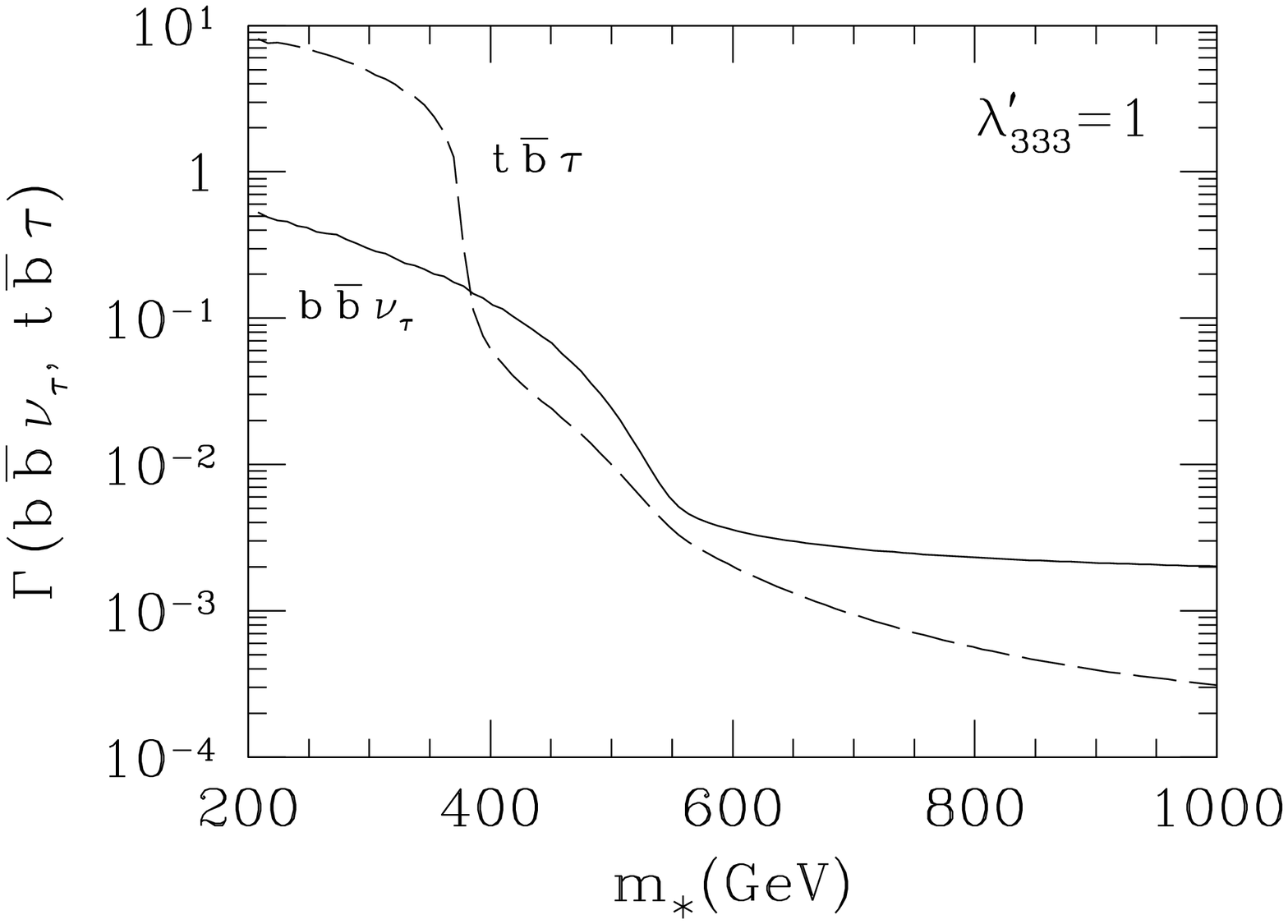}
\includegraphics*[scale=0.35]{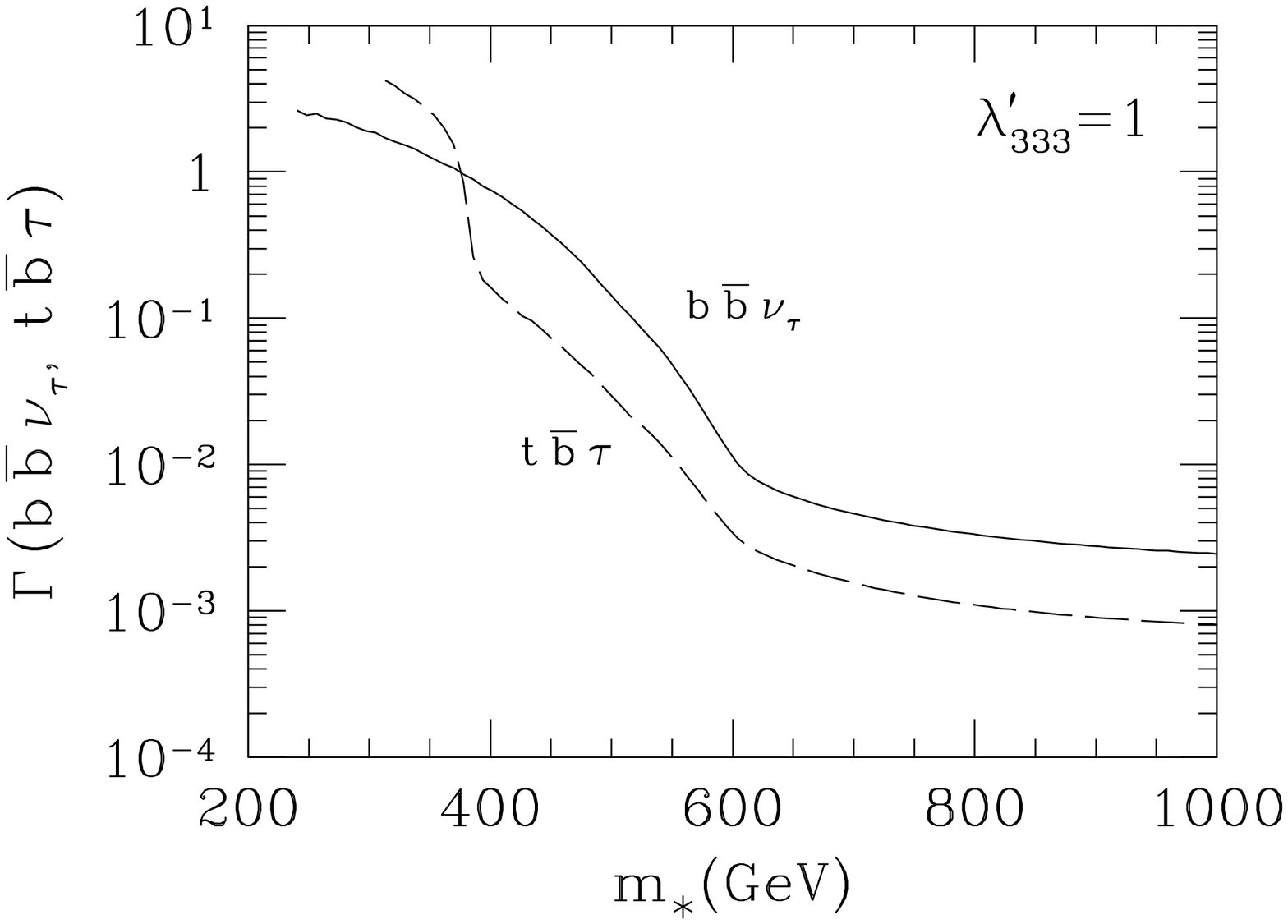}
\caption{Solid lines show  the $\tilde \chi_1^0$ decay 
width for the $b \bar b \nu_\tau$  channel  and dashed lines for 
the $ t \bar b \nu$ channel.  The left panel is for $\tan \beta = 3$ with 
the trilinear soft term being $350$ GeV whereas in the  right panel
$\tgb = 30$ and the  trilinear soft term is $150$  GeV. $m_{\tilde B }$, 
slepton mass and $\mu$ are all taken to be 600 GeV.}
\label{fig4}
\end{figure}
Figure~\ref{fig4} shows the $b \bar b \nu_\tau,  t \bar b \nu$ decay widths,
as a function of  the squark mass scale $m_*$. The massive mode dominates 
at lower $\tan \beta$ and the massless one for the higher values. Thus, to 
summarize, we find that with $\lambda'_{333} $ dominant, the massive decay 
has a large width for low $\tan \beta$ and large higgsino--gaugino mixing. 
Otherwise the width for the massless mode is larger, though that for 
massive mode is also non-negligible.
All the plots have been shown for $\lambda'_{333} = 1$. In the resonant 
sfermion region one finds very little dependence of the width on 
$\lambda'_{333}$ and otherwise the widths for other values of the 
couplings simply scale down like $\lambda^{'2}_{333}$.

\begin{figure}[hbt]
\includegraphics*[scale=0.38]{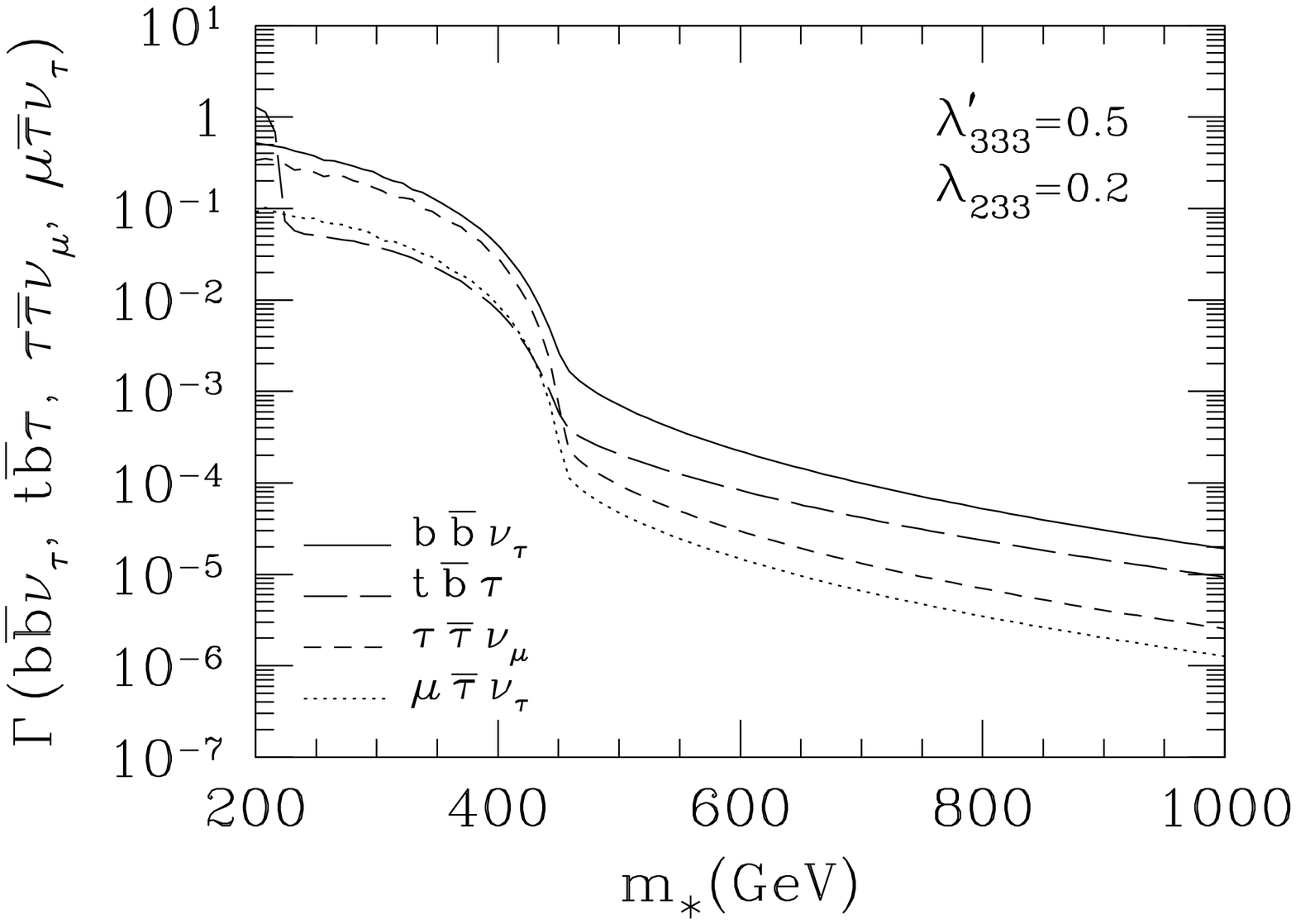}
\includegraphics*[scale=0.38]{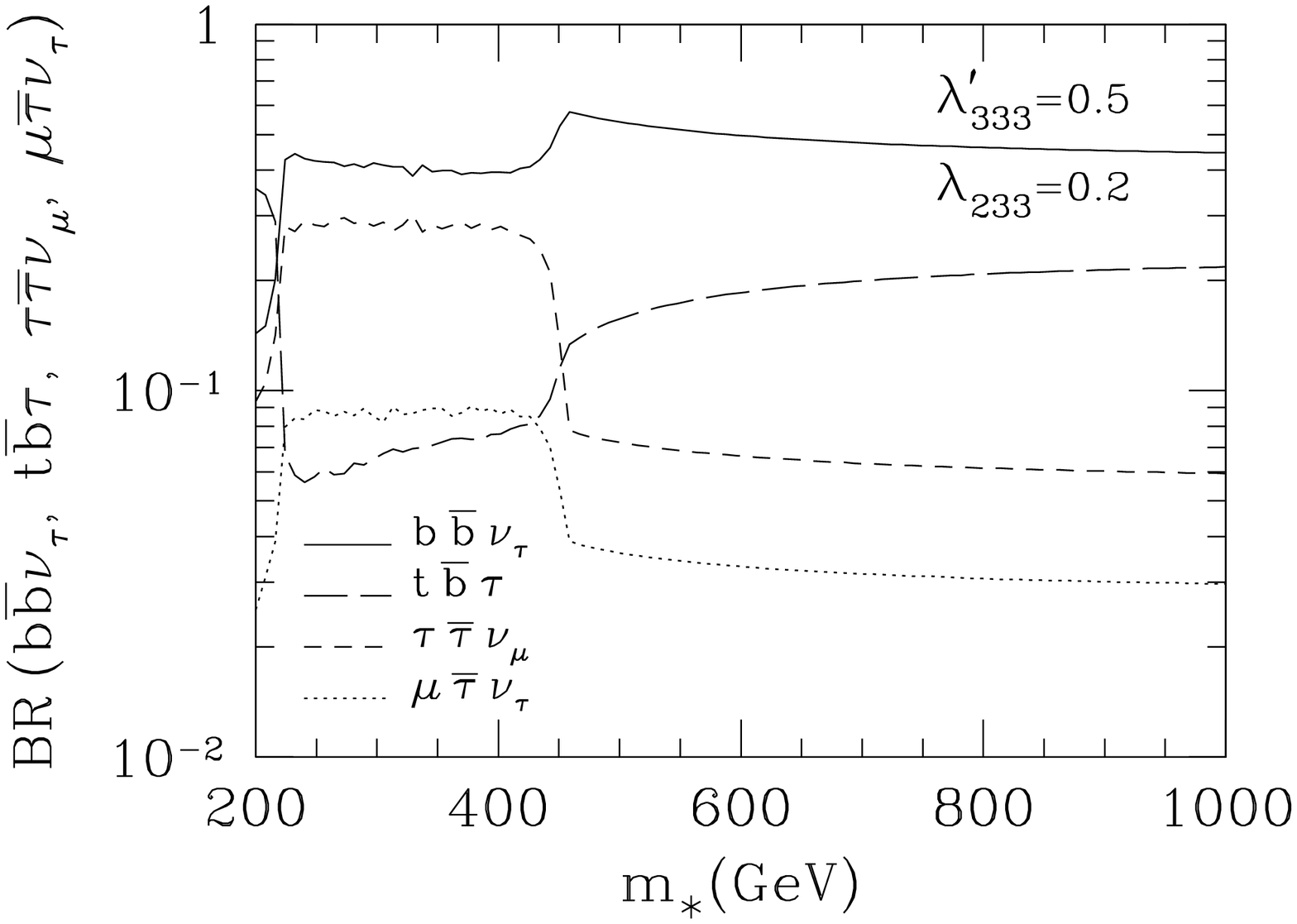}
\caption{Left panel shows the decay widths of $\tilde \chi_1^0$ for  
$\lambda`_{333} = 0.5, \lambda_{233} = 0.2$ for various final states 
as indicated in the figure. The right panel shows branching ratios for
the same. These are for no $L$--$R$ mixing, $\mu = 500$ GeV, 
$\tgb = 3$ and $m_{\tilde B} = 500$ GeV.}
\label{fig5}
\end{figure}
The analysis can be easily generalized to the case where there are more
than one large $\rp$ couplings, and the results
are shown in Figure~\ref{fig5}. We notice that the widths into different
final states are not too dissimilar, and more importantly the branching
ratio into a final state with $t$ is not too small.

\section*{Conclusions}
It is important to have a collider 
probe of the $\rp\ \lambda$ and $\lambda'$ couplings with more than one 
third-generation index. Further, single charged slepton production via 
the $\rp$ coupling $\lambda'_{i33}$  can produce final states, which can be 
confused with the $H^\pm$ signal. Even for smaller but dominant 
$\rp\  \lambda$ and $\lambda'$ couplings with more than one third-generation 
index, the three-body $\rp$ decays of $\tilde \chi_1^0, \tilde \chi_1^+$
can have important phenomenological consequences for new particle
searches at the future colliders. If the $\tilde \chi_1^0$ is `Wino'-like,
its decay widths increase by over an order of magnitude. If the 
$\tilde \chi_1^0$ is lighter than the $t$, its $\rp$ decays will be 
dominantly into massless fermions: $b \bar b \nu_\tau$ for dominant 
$\lambda'_{333}$ coupling, $c \bar b \tau, s \bar b \nu_\tau$  and 
the charge conjugate modes $\bar c b \bar\tau, \bar s b {\bar\nu}_\tau$  
for dominant $\lambda'_{323}$, and for $\lambda_{233}$ into 
$\bar \mu \tau \nu_\tau$ and $\mu \bar\tau {\bar\nu}_\tau$. For 
a $\tilde \chi_1^0$ heavier than the $t$, the massive
decay modes can become competitive for large $L$--$R$ sfermion mass term
and/or for substantial mixing in the higgsino/gaugino sector, at not too
large $\tan \beta$. All the $\rp$ decays of the $\tilde \chi_1^0$ produce
final states with like-sign dileptons as the tell-tale
signature of the $\LV$, as long as the final state $t$ decays
hadronically.

\noindent{\bf Acknowledgements} \\
I wish to thank my collaborators F. Borzumati, J.L. Kneur and F. Takayama 
for a nice collaboration, which resulted in the publication on which this talk 
is based.  

\end{document}